\documentclass[aps,pre,11pt,onecolumn,notitlepage,superscriptaddress,floatfix]{revtex4-2}

\usepackage{epsfig}
\usepackage{amssymb,amsmath}

\usepackage[mathlines]{lineno}

\usepackage{etoolbox} 

\newcommand*\linenomathpatchAMS[1]{%
  \expandafter\pretocmd\csname #1\endcsname {\linenomathAMS}{}{}%
  \expandafter\pretocmd\csname #1*\endcsname{\linenomathAMS}{}{}%
  \expandafter\apptocmd\csname end#1\endcsname {\endlinenomath}{}{}%
  \expandafter\apptocmd\csname end#1*\endcsname{\endlinenomath}{}{}%
}

\newcommand\Tstrut{\rule{0pt}{2.9ex}}         

\expandafter\ifx\linenomath\linenomathWithnumbers
  \let\linenomathAMS\linenomathWithnumbers
  \patchcmd\linenomathAMS{\advance\postdisplaypenalty\linenopenalty}{}{}{}
\else
  \let\linenomathAMS\linenomathNonumbers
\fi

\usepackage{array}
\usepackage{graphicx,color,subfigure,bm,amsmath,amssymb}
\usepackage{hyperref}
\usepackage[mathlines]{lineno}
\hypersetup{
    colorlinks=true,
    linkcolor=cyan,
    filecolor=magenta,      
    urlcolor=blue,
    citecolor=blue,
}
\usepackage{graphicx}
\usepackage{dcolumn}
\usepackage{bm}
\usepackage{color}

\newcommand{\lam}{\lambda}

\newsavebox{\astrutbox}
\sbox{\astrutbox}{\rule[-5pt]{0pt}{20pt}}

\newcommand{\sst}[1]{\scriptscriptstyle{#1}}

\newcommand{\Nab}{\mbox{$\bm \nabla$}}

\newcommand{\vecx}{\mbox{$\bm x$}}

\newcommand{\vecxi}{\mbox{$\bm \xi$}}

\newcommand{\vecu}{\mbox{$\bm u$}}

\newcommand{\vecv}{\mbox{$\bm v$}}
\newcommand{\vecr}{\mbox{$\bm r$}}

\newcommand{\vecI}{\mbox{$\bm I$}}

\newcommand{\tlam}{\tilde{\lambda}_{\sst 0}}
\newcommand{\Gam}{\Gamma_{\sst 1}}

\newcommand{\Omz}{\Omega_{\sst 0}}

\newcommand{\vecsig}{\mbox{\boldmath$\sigma$}}

\newcommand{\icomp}{\mbox{i}}
\newcommand{\icomps}{\mbox{\scriptsize i}}
\newcommand{\expo}{\mbox{e}}

\let\oldequation\equation
\let\oldendequation\endequation

\renewenvironment{equation}
  {\linenomathNonumbers\oldequation}
  {\oldendequation\endlinenomath}
  
\begin{document}

\title{Direct Measurement of Unsteady Microscale Stokes Flow\\ Using Optically Driven Microspheres}

\author{Nicolas Bruot}
\affiliation{Cavendish Laboratory and Nanoscience Centre, University of Cambridge, Cambridge 
CB3 0HE, United Kingdom}
\affiliation{Institut de Physique de Nice, CNRS, UMR No. 7010, Universit\'e C\^ote 
d'Azur, 06108 Nice, France}
\author{Pietro Cicuta}
\affiliation{Cavendish Laboratory and Nanoscience Centre, University of Cambridge, Cambridge 
CB3 0HE, United Kingdom}
\author{Hermes Bloomfield-Gad\^{e}lha}
\affiliation{Department of Engineering Mathematics \&
Bristol Robotics Laboratory,University of Bristol, Bristol BS8 1UB, United Kingdom}
\author{Raymond E. Goldstein}
\email[For correspondence: ]{R.E.Goldstein@damtp.cam.ac.uk}
\affiliation{Department of Applied Mathematics and Theoretical Physics, Centre for Mathematical
Sciences, University of Cambridge, Cambridge CB3 0WA, United Kingdom}
\author{Jurij Kotar}
\affiliation{Cavendish Laboratory and Nanoscience Centre, University of Cambridge, Cambridge 
CB3 0HE, United Kingdom}
\author{Eric Lauga}
\affiliation{Department of Applied Mathematics and Theoretical Physics, Centre for Mathematical
Sciences, University of Cambridge, Cambridge CB3 0WA, United Kingdom}
\author{Fran\c cois Nadal}
\email[For correspondence: ]{F.R.Nadal@lboro.ac.uk}
\affiliation{Wolfson School of Mechanical, Electrical and Manufacturing Engineering, Loughborough
University, Loughborough LE11 3TU, United Kingdom}

\date{\today}

\begin{abstract}

A growing body of work on the dynamics of eukaryotic flagella has noted that
their oscillation frequencies are sufficiently high that the viscous penetration depth
of unsteady Stokes flow is comparable to the scales over which flagella 
synchronize.  Incorporating these effects into theories of synchronization requires an understanding
of the global unsteady flows around oscillating bodies.
Yet, there has been no precise experimental test on the microscale of the most basic aspects of
such unsteady Stokes flow: the orbits of passive tracers and the position-dependent phase lag between the oscillating 
response of the fluid at a distant point and that of the driving particle.
Here, we report the first such direct Lagrangian measurement of this unsteady flow. The method uses 
an array of $30$ submicron tracer particles positioned by 
a time-shared optical trap at a range of distances and angular positions with respect to a 
larger, central particle, 
which is then driven by an oscillating optical trap at frequencies up to $400$ Hz.  In this 
microscale regime, the tracer dynamics is considerably simplified by the smallness of both 
inertial effects on particle motion and finite-frequency corrections to the Stokes drag law.
The tracers are found to display elliptical Lissajous figures whose orientation and geometry are 
in agreement with a low-frequency expansion of the underlying dynamics, 
and the experimental phase shift between motion parallel and orthogonal to the oscillation axis 
exhibits a predicted scaling form in distance and angle.  Possible implications of these results for
synchronization dynamics are discussed.

\end{abstract}

\maketitle

\section{Introduction}

In his landmark 1851 paper on viscous fluids \cite{Stokes1851}, George Gabriel Stokes not only
developed the theoretical framework for understanding the competition between inertial and viscous
forces, but he also considered several physical situations in which that competition is particularly simple to
analyze.  These include his celebrated problems I and II \textemdash~viscous fluid in the half space adjacent
to a no-slip wall that is impulsively started into motion or oscillated from side to side at some
frequency $\omega$ \textemdash~ 
as well as the case of a sphere oscillated back and forth.  From these oscillatory cases and his newly 
identified `index of friction' (what we now term the kinematic viscosity $\nu=\eta/\rho_f$, $\eta$ and $\rho_f$ being the
dynamic viscosity and density of the fluid), he identified from the diffusion equation $u_t=\nu\,u_{xx}$, for a component $u$
of the fluid velocity, the viscous penetration length 
\begin{equation}
    \delta=\left(2\nu/\omega\right)^{1/2}
    \label{delta_define}
\end{equation}
as the distance over which oscillatory motions decay away from the driving surface.  Furthermore, the fluid 
oscillations at some distance $r$ from the driving body are phase shifted relative to the drive by an 
angle proportional to $r/\delta$. 

There is, of course, no doubt of the {\it validity} of his analysis of these particular problems.
Yet, in the motivating biophysical context we consider here there has been longstanding uncertainty about 
the {\it relevance} of unsteadiness to phenomena that are strongly in the Stokes regime, 
such as the beating of eukaryotic flagella and the motion of 
tracer particles in flows driven by flagellated organisms.  For example, in models for
ciliate propulsion \cite{Brennen}, it has been recognized that around large microorganisms 
covered in a dense cilia carpet, unsteady effects are significant within a region near the organism 
surface of width 
$\sim\!\delta$, and outside the flow may be 
considered steady.  Consider for example the well-studied multicellular organism alga {\it Volvox} \cite{ARFM}, a spheroid 
of radius $\sim\!200\, \mu$m, covered with thousands of biflagellated somatic cells each $10\,\mu$m in diameter, 
spaced some $20\,\mu$m apart, 
whose flagella of length $\ell\sim 25\, \mu$m beat at a frequency $f\sim 25\,$Hz.  The
viscous penetration depth $\delta\!\sim\! 110\, \mu$m is significantly less than the
organism's circumference, but it is intriguingly close to the wavelength of metachronal waves that
{\it Volvox} exhibits \cite{Volvox_metachronal1,Volvox_metachronal2}.  These are 
long-wavelength phase modulations of the beating in the form like that of a 
stadium wave which, in {\it Volvox}, have a wavelength $\sim\! 100\,\mu$m \cite{Volvox_metachronal2}.
Thus, even nearest-neighbor somatic cells have a significant phase shift.  This is in contrast to the 
situation in {\it Chlamydomonas}, the unicellular relative of \textit{Volvox} whose size is comparable to {\it Volvox}
somatic cell and whose two $10-12\,\mu$m flagella 
are mounted just a few microns apart and beat at $\sim 50\,$Hz; the phase shift between the
flagella is indeed rather small. 
Yet, nearly all models of flagellar synchronization and in particular of 
metachronal wave formation \cite{MW1,MW2,MW3,MW4,MW5,MW6,MW7,MW8,MW9} assume as a starting point the 
{\it steady} Stokes equation.  
It is only recently, in the context
of the dynamics of tracer particles in flows generated by the beating flagella of alga \cite{Leptos,Tam,TamJFM,TamEIF} 
that unsteadiness has been identified as a potentially significant feature of biophysical flows. 

With the goal of motivating further studies of these phenomena, we introduce an 
experimental setup by which optical trapping
methods \cite{BruotCicuta,Cicuta} are used to measure the motion induced by unsteadiness over 
a broad angular
sector around a central oscillated microsphere.  This setup allows for a precise test of the underlying 
microhydrodynamic theory, with results that are complementary to recent 
experimental studies of oscillatory flows driven by the more complex beating of {\it Chlamydomonas} flagella, where the
phase lag experienced by tracer particles was measured directly \cite{Tam,TamEIF}.  As shown in earlier work 
on synchronization \cite{hydrosynchro}, 
the far-field flows due to eukaryotic flagella are accurately represented by moving
point forces.  Thus, we expect the present results to inform future analysis of flagellar interactions 
on the basis of simplified representations of their dynamics. 

We begin in Sec. \ref{sec:methods} with a description of the experimental setup, the frequency 
response of the optical
trap used to oscillate a microsphere surrounded by an array of tracers, and a discussion of
inertial corrections to the motion of the microspheres.  
The results presented in Sec. \ref{sec:results} comprise the motion of 
tracer particles at varying
distances and angular positions relative to the driven microsphere. The theoretical analysis of their 
Lagrangian dynamics is done with the Eulerian flow field of the classical solution for motion around an oscillating sphere.
A low-frequency expansion, valid when $\delta$ is large compared to distances from the 
central sphere, is used to obtain a geometrically simple result for the tracer trajectories, which are elliptical 
Lissajous figures. 
As the tracers are submicron, they exhibit substantial
thermal fluctuations which compete with the deterministic displacements from the oscillating 
flow, and this can be quantified by a suitable P{\'e}clet number that varies with oscillation 
frequency and distance from the central particle.
The phase shift between motion along the two Cartesian directions, which is 
responsible for the shape of those orbits, is calculated in the low-frequency limit and found to be in excellent agreement with the data.
The implications of the observed
phase shift between the driven and tracer particles on synchronization processes relevant to 
the biomechanics of cilia are discussed in the concluding Section \ref{sec:discuss_concl}.

\section{Setup and experimental methods \label{sec:methods}}

\begin{figure}[t]
\begin{center}
\scalebox{0.95}{\includegraphics{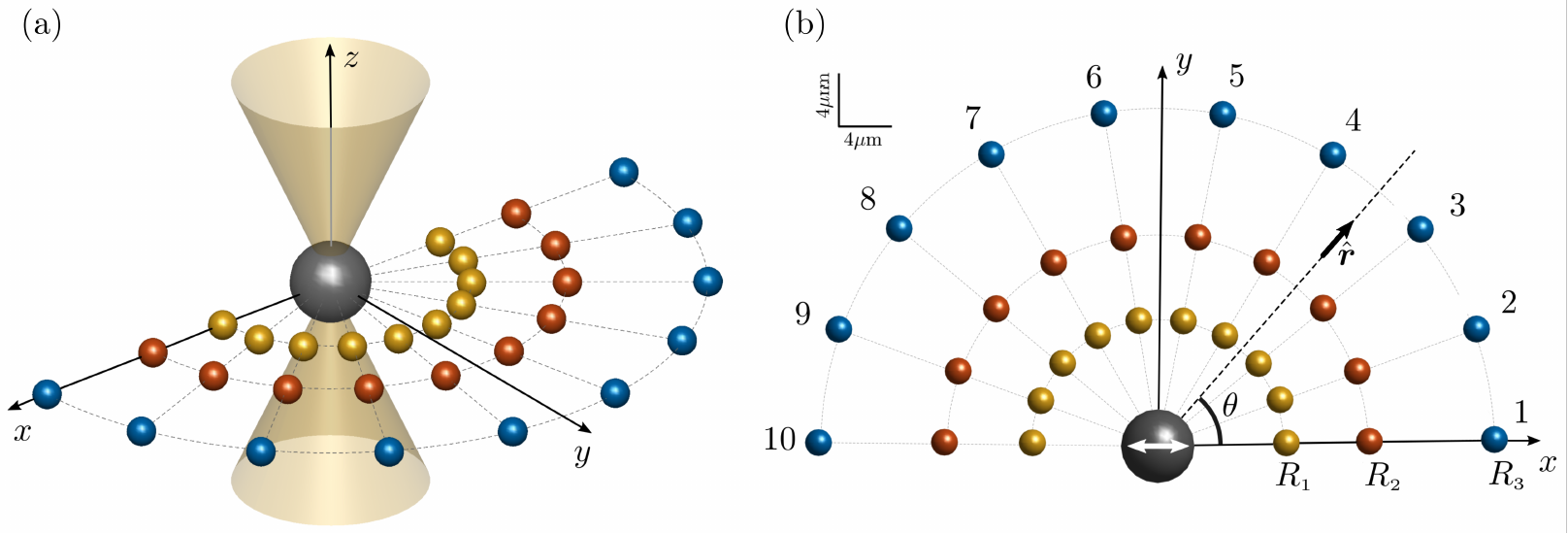}}
\caption{Schematic of the experiment. (a) The large silica microsphere (grey) of radius 
$a_{\sst 0} = 2.77$\,$\mu$m 
is oscillated at various frequencies by means of an optical trap. Previous to the actuation of the 
large particle, passive microspheres (yellow, red, blue) 
with radius $a_{\sst 1}= 0.505$\,$\mu$m are placed at three different 
distances $R_{1} = 9.5118$\,$\mu$m, $R_{2} = 15.853$\,$\mu$m and $R_{3} = 25.365$\,$\mu$m from the center 
of the driving bead. They are held in position {\it via} a multiple trap system, 
and released automatically upon actuation of the central sphere, whose motion 
along the $x$-axis is shown by a double white arrow.\label{fig:geometry}}
\end{center}
\end{figure}

A large silica microsphere (radius $a_{\sst 0} \sim 2.77$ $\mu$m, $\rho_{\sst 0} = 2.65\times 10^3$\,kg\,m$^{-3}$) 
is forced to oscillate horizontally along the $x$-axis with an amplitude 
$\xi_{\sst 0}(t) = \xi_{\sst 0}\,\exp(\icomp \omega t)$ 
in water (density $\rho_w = 10^3$~kg~m$^{-3}$,
viscosity $\eta = 10^{-3}$~Pa\,s) by means of optical tweezers.
Smaller passive polystyrene microspheres (radius $a_{\sst 1} = 0.505$ $\mu$m, 
density $\rho_{\sst 1} = 1.05\times 10^3$\,kg\,m$^{-3}$), also referred to as {\it probes} or {\it tracers},
are located in the horizontal $(x,y)$- plane at three different distances $R_i$ ($R_{1} = 9.5118$ $\mu$m,
$R_{2} = 15.853$ $\mu$m and $R_{3} = 25.365$ $\mu$m) from the central sphere, 
and ten different angles $\theta_j$ ($j=1\cdots 10$) equally spaced
within the interval $[0,\pi]$, as shown in Fig.~\ref{fig:geometry}. The Lagrangian displacement of a polystyrene sphere
located at $(R_i,\,\theta_j)$ due to the flow generated by the central bead is denoted by 
$\vecxi^{\sst{ij}} = (\xi_x^{\sst{ij}},\xi_y^{\sst{ij}})$. 

The particles used in each experimental run were extracted from dilute suspensions of silica (Bangs Laboratories) and 
polystyrene (Polyscience) microspheres.  The polystyrene and silica beads 
were sufficiently dilute that no particles other than those used as
oscillator or probes interfered with the laser beam during an experiment. The solution was sealed between a microscope
slide and a coverslip separated by a 150~$\mu$m gap and held together by NOA 68 UV-cured glue.  Microspheres
were trapped at least $50$~$\mu$m from the chamber walls to minimize any wall-particle interactions. 

The tweezers setup is as described elsewhere \cite{Kotar2013,MW9,Maestro2018}. Briefly, 
the beam of a diode-pumped solid-state laser (CrystaLaser IRCL-2W-1064, 1064~nm wavelength, 2~W maximum output power) 
is deflected by a pair of acousto-optic deﬂectors (AA.DTS.XY-250@1064\,nm, AA Opto-Electronic) and directed to the 
back illumination port of a Nikon Ti-E inverted microscope, in which the beam is reflected by a dichroic mirror and 
focused on the sample by a Nikon Plan Apo VC 60x objective (NA = 1.20). 
The samples were viewed with brightfield illumination and the dynamical response of the microspheres to the 
oscillating flow was recorded by high speed camera (Phantom V5.1) at 25,000 frames per second. The acousto-optic 
deflector allows for time-sharing of the laser beam so that multiple traps located at prescribed positions can be 
created in the $(x,y)$ focal plane. 
The stiffness of the trap for the silica  when it is trapped alone was determined to be $k = 50 \pm 1\,$pN~$\mu$m$^{-1}$, by 
measuring the standard deviation of the particle's thermal fluctuations.

Initially, the silica bead is trapped at the origin of the coordinate system 
while the ten polystyrene particles are held at locations  $(R_i,\theta_j)$, one $R_i$ at a time. The central particle
is driven by moving its optical trap along the $x$-axis by sampling 
the path as $N_p$ points which are cyclically visited by 
the acousto-optic deflector. Once the  polystyrene particles are released, the entire laser power
is reassigned to the oscillating trap and the 
trajectories of the silica and polystyrene particles are recorded.  The corresponding tracks in the 
$(x,y)$-plane are 
extracted using a bespoke image segmentation tracking algorithm.
The system was optimized to reach driving frequencies up to $400$~Hz for a trap oscillation amplitude of 
$2.15\,\mu$m.  The oscillation frequency and amplitude are limited by the dynamics of the particle 
(see below) and the finite size of the optical potential well. Once they are optimized, the maximal 
$N_p$ for path sampling is obtained from the highest target frequency
and a time-sharing frequency of 20\,kHz set by the bespoke electronics that control the acousto-optic deflector;
$N_p = 50$ in experiments reported here.  Under such conditions, the drive bead 
follows a sinusoidal pattern, even when it does not remain close to the trap center. 

\begin{table*}
\caption{\label{tab:table1}Physical quantities for experiments in water, in a convenient system of units.}
\begin{ruledtabular}
\begin{tabular}{cccc}
Symbol&Definition&Quantity&Value\\ 
\hline \Tstrut
$\nu$ & $\eta/\rho_w$ & kinematic viscosity of water & $10^{6}$~$\mu$m$^2$~s$^{-1}$\\
$\delta$ & $(2\nu/\omega)^{1/2}$ & viscous penetration depth & 28-80 $\mu$m\\
$a_{\sst 0}$& & radius of driven microsphere & $2.77$ $\mu$m\\
$a_{\sst 1}$& & radius of tracer microspheres& $0.505$ $\mu$m\\
$D_{\sst 1}$& k$_B$T/6$\pi\eta a_1$ & diffusion constant tracer microspheres& $0.43$ $\mu$m$^2$~s$^{-1}$\\
$\rho_{\sst 0}$& & density of driven microsphere & $2.65\times 10^3$\,kg\,m$^{-3}$\\
$\rho_{\sst 1}$& & density of tracer microspheres & $1.05\times 10^3$\,kg\,m$^{-3}$\\
$R_i$& $i=1,2,3$ & unperturbed radial distances of tracers & $9.5, 15.8, 25.4$\,$\mu$m\\
$\theta_j$ & $(j-1)\pi/9$ & unperturbed angular position of tracers & $j=1,\ldots,10$\\
$k$& & optical trap stiffness & $50\,$pN~$\mu$m$^{-1}$\\
$\omega$ & $2\pi f$& oscillation frequency of driven microsphere & $2\pi \times$ ($50-400$ Hz)\\
$\zeta_{\rm trap}$ & & oscillation amplitude of optical trap & $2.15\,\mu$m\\
$\xi_{\sst 0}(\omega)$ & $\vert \zeta_{\sst 0}(\omega)\vert$& oscillation amplitude of driven microsphere & ($0.4-1$)$\times\xi_{\rm trap}$\\
${\bf \xi}^{ij}$& & Lagrangian displacement of sphere at ($R_i$,$\theta_j$)& $<0.4\times\xi_{\sst 0}$\\
$\chi^{ij}$& max($\xi^{ij}_x/\xi_{\sst 0}$) & scaled maximum Lagrangian $x$-displacement of sphere at ($R_i$,$\theta_j$)& $< 0.4$\\
$\phi$& $\phi_y-\phi_x$& relative phase lag of $x$- and $y$-components of tracers& $\lesssim \pi/2$ \\
\end{tabular}
\end{ruledtabular}
\end{table*}

The dynamics of a microsphere forced by an optical trap whose position is laterally oscillated is a 
well-studied problem \cite{trapping}.
When, as is the case here, the displacement of the driven particle is sufficiently small that the 
optical force exerted on the sphere is linearly proportional to the distance from the axis of the beam, 
and the trap and particle positions oscillate as $\zeta_{\sst 0}\expo^{i\omega t}$ 
and $\zeta_{\rm trap}\expo^{i\omega t}$, then momentum conservation in the $x$-direction takes the form
\begin{equation}
\left(-\rho_{\sst 0}\mathcal{V}_{\sst 0}\omega^2 + 6\pi\eta a_{\sst 0}\,
\Omz\,\icomp\omega\right)\zeta_{\sst 0} = -k\left(\zeta_{\sst 0}-\zeta_{\sst trap}\right) 
,
\label{eq:mom_trap}
\end{equation}
where the left hand side represents the inertia of the particle itself (whose volume is $\mathcal{V}_{\sst 0} 
= 4\pi a_{\sst 0}^3/3$) and the drag force, while the right hand side is the trap force.
The drag force, found in the original derivation by Stokes in 1851 \cite{Stokes1851} and
in more modern treatments \cite{MazurBedeaux,KimKarrila} has a factor $\Omz$ that corrects the
familiar zero Reynolds number Stokes drag for fluid inertia,
\begin{equation}
\Omz = 1 + \alpha_{\sst 0} + \frac{\alpha_{\sst 0}^2}{9}\;\;\;\mbox{with}\;\;\alpha_{\sst 0} = 
(1+i)\frac{a_{\sst 0}}{\delta}.
\label{eq:omegazero}
\end{equation}
From \eqref{eq:mom_trap} and \eqref{eq:omegazero}, we identify three characteristic times scales associated 
with the experiment, the shortest of which is that for fluid momentum to diffuse on the scale of particle,
\begin{equation}
    \tau_{d}=\frac{a_{\sst 0}^2}{\nu}\sim 8\times 10^{-6} {\rm s}\,.
    \label{taud}
\end{equation}
Next is that for inertial oscillations of the sphere in the trap,
\begin{equation}
    \tau_i=\left(\frac{\rho_{\sst 0}\mathcal{V}_{\sst 0}}{k}\right)^{1/2}\sim 7\times 10^{-5} {\rm s}\,,
\end{equation}
and finally the time scale over which a particle viscously relaxes to the trap center,
\begin{equation}
    \tau_r=\frac{6\pi\eta a_{\sst 0}}{k}\sim 10^{-3} {\rm s}~.
\end{equation}
  
\begin{figure}[t]
\centering
\scalebox{0.98}{\includegraphics{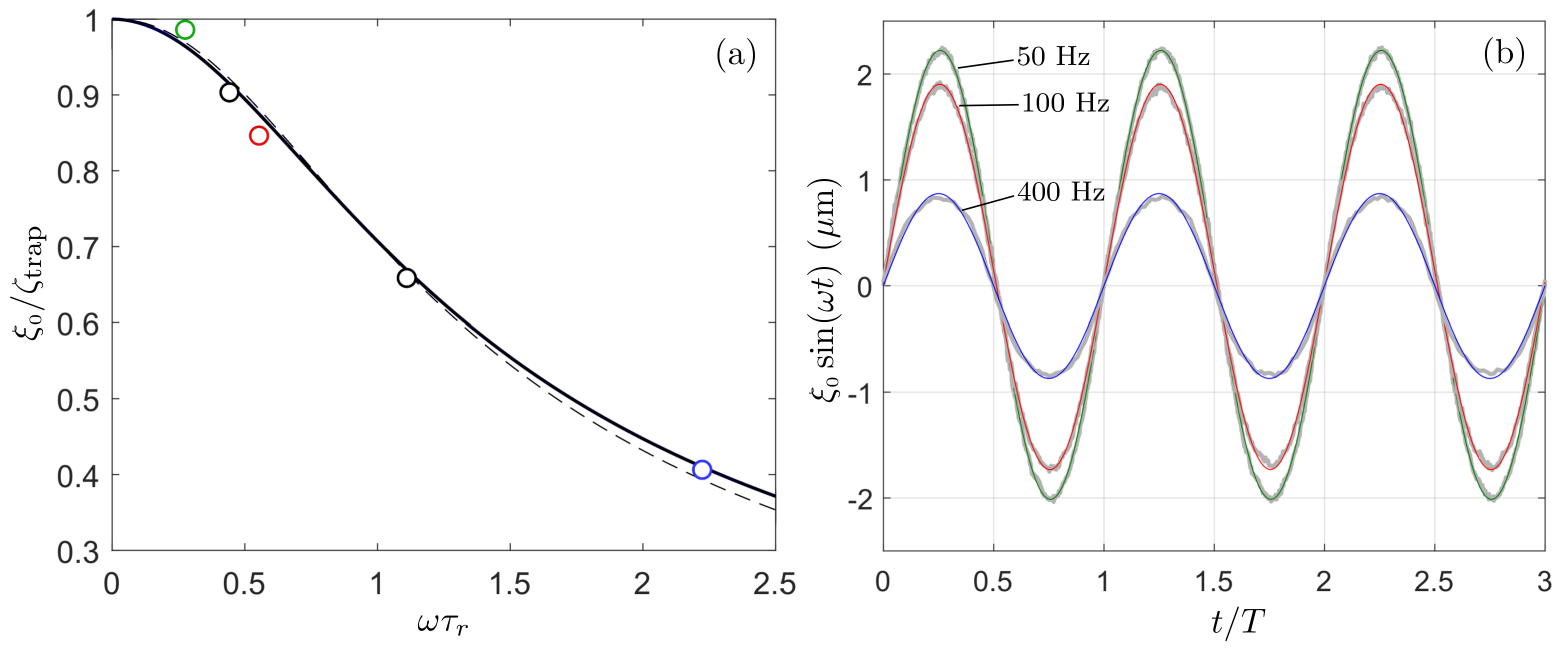}}
\caption{Oscillations of the driven microsphere.  (a) Amplitude of oscillations relative to trap oscillation 
amplitude as a function of frequency. The theoretical prediction in Eq. \ref{eq:ampl_central_bead} with
$\tau_d=0$ (solid blue line) matches well the experimental data
(open circles). Dashed red line indicates small correction obtain by including finite $\tau_d$ corrections. 
(b) Measured $x$-position of the bead as a function of the time (thick gray lines),
fitted with a sinusoidal function (colored lines) at three frequencies. For clarity of presentation, the origin of 
time has been shifted to align the different curves.
The good match between the experimental profiles and the fitting functions
validates the linear approach used to derive Eq.~(\ref{eq:ampl_central_bead}).\label{fig:osc_central_bead}}
\end{figure}

With these definitions, we have
\begin{equation}
    \zeta_{\sst 0}=\frac{\zeta_{\rm trap}}{1+i\omega\tau_r\Omz-\left(\omega \tau_i\right)^2}\,.
    \label{eq:zeta}
\end{equation}
At the highest frequencies probed ($400$ Hz), $\omega\tau_i\sim 0.18$ and 
particle inertia contributes only a few percent to the response of microspheres.
Similarly,
from \eqref{eq:omegazero} we deduce that the maximum contribution of momentum diffusion has $\omega\tau_d\sim 0.02$,
so $\vert \alpha_{\sst 0} \vert\sim 0.14$, yielding a modest correction to the force amplitude $\Omz$, 
and higher-order contributions are negligible.  
By the quadratic scaling of $\tau_d$ with sphere radius, this simplification is due to the
use of microspheres.  The complex structure of \eqref{eq:zeta} implies that there is a phase shift between the 
trap and the driven particle, but as we are interested in the response of the tracers 
relative to the driven microsphere, we ignore that phase shift and use the motion of the driven bead as the
time reference in the following and define $\xi_{\sst 0}(\omega)=\vert\zeta_{\sst 0}\vert$;  
we adopt a time origin such that the driven particle's
position is $\xi_{\sst 0}\cos(\omega t)$.
Neglecting particle inertia and quadratic terms in
$\Omz$, we have
\begin{equation}
\xi_{\sst 0}(\omega)\simeq \zeta_{\rm trap}\left[\left(1-(\omega\tau_r)(\omega\tau_d/2)^{1/2}\right)^2 
+ \left(\omega\tau_r\right)^2\left(1+(\omega\tau_d/2)^{1/2}\right)^2\right]^{-1/2},
\label{eq:ampl_central_bead}
\end{equation}
which represents only minor deviations from the Lorentzian form $[1+(\omega\tau_r)^2]^{-1/2}$. This ratio and
its Lorentzian approximation are plotted in Fig.~\ref{fig:osc_central_bead}\,a as a function of the 
rescaled frequency $\omega\tau_r$. With no free parameters the agreement between the prediction of 
\eqref{eq:ampl_central_bead} and experiment is excellent. Figure \ref{fig:osc_central_bead}b shows the 
accurately sinusoidal displacement of the oscillated particle. 

\section{Results \label{sec:results}}
\begin{figure}[t]
\begin{center}
\vspace{0cm}
\scalebox{0.97}{\includegraphics{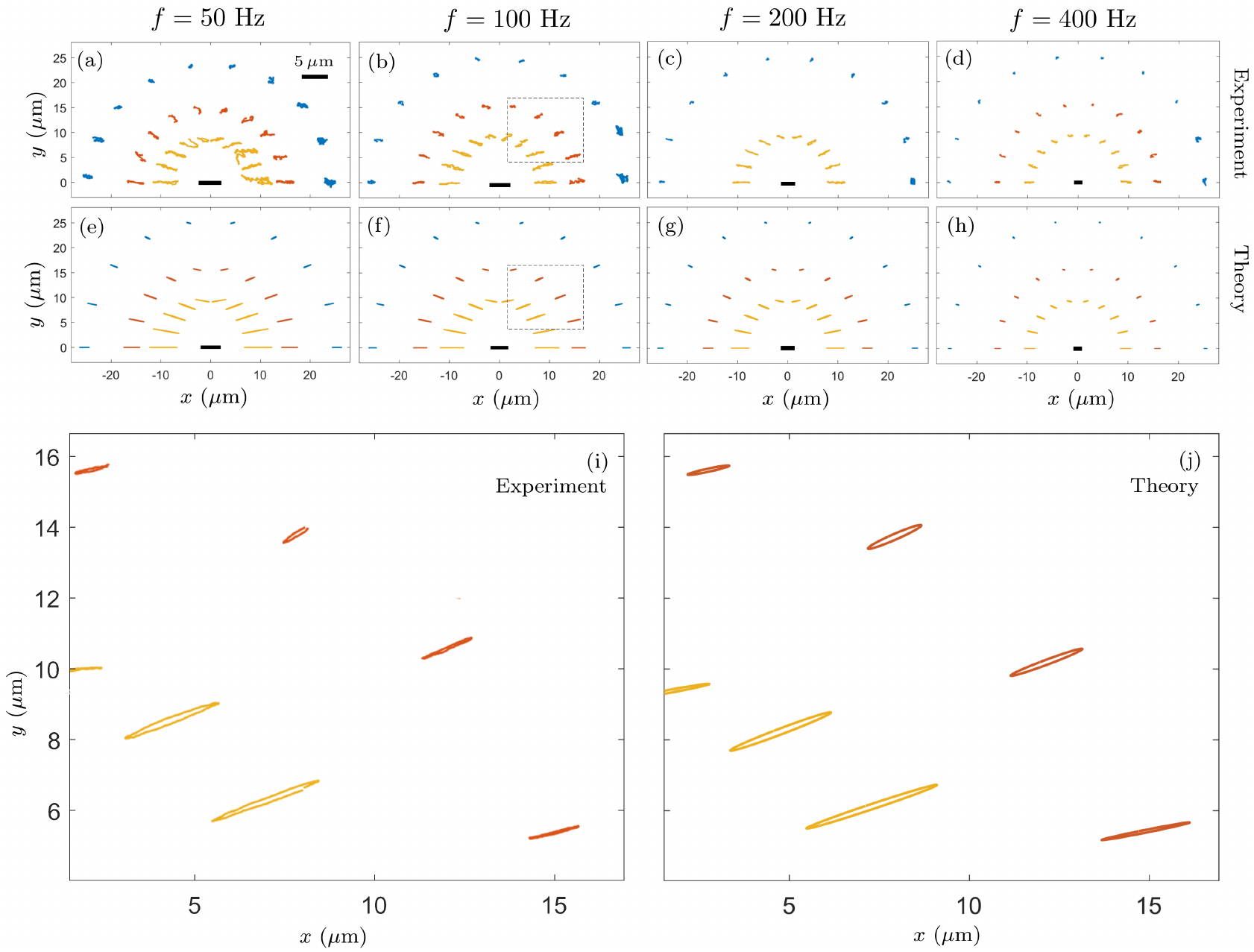}}
\caption{Oscillatory tracer dynamics with tracer displacements magnified by a factor $4$, 
while their equilibrium locations are to scale. (a-d) Tracer trajectories over one 
period. Colors 
(yellow, red,blue) indicate different radial distances, as in Fig. \ref{fig:geometry}. The peak-to-peak 
displacement of the central silica bead is indicated by the heavy black line. (e-h) Theoretical trajectories
using approach given in text. 
(i-j) Enlarged view of trajectories in the boxed region shown in (b,h).  (i) shows average cycles from experiments while (j) is simply a zoomed representation of (f). The flat elliptical trajectories, 
which arise from the phase shift between the components $\xi_x^{\sst{ij}}$ and $\xi_y^{\sst ij}$ of 
the displacement, are accurately captured by the theory.
\label{fig:traj}}
\end{center}
\end{figure}

Raw trajectories of the probes during a single period of oscillation are shown in 
Figs.~\ref{fig:traj}\,a-d, where for clarity, the probe displacements are magnified by a factor of $4$, 
while their mean positions are to scale \cite{missingdata}.  
Each of the trajectories is an elliptical Lissajous figure whose major and minor axes vary
systematically with angular position $\theta$ and distance $R$ from the driven bead. 
It is also clear that the probe trajectories have a degree of stochasticity superimposed on their background
motion.

A quantitative treatment of the probe trajectories and an assessment of the importance of 
Brownian motion begin with the unsteady velocity field $\vecu(\vecr,t)$ due to a sphere oscillating with velocity 
$v_{\sst 0}\,\expo^{\icomps\omega t}\,\hat{\vecx}$.
That velocity field satisfies the full Navier-Stokes equations which, if made dimensionless by the time $1/\omega$, 
a length $L$ and velocity $U$ has two Reynolds numbers \cite{KimKarrila},
\begin{equation}
\frac{L^2\omega}{\nu}\vecu_{t'}' + \frac{UL}{\nu}\vecu'\cdot\Nab'\vecu' = \Nab'\cdot\vecsig',
\label{eq:NS_adim}
\end{equation}
where $\vecsig'$ is the non-dimensional hydrodynamic stress. If $L$ is on the scale
of the sphere radius $a_0$, then $U\sim a_0\omega$ and $L^2\omega/\nu$ and $UL/\nu$ are both 
$\sim\! a_0^2\omega/\nu= \vert \alpha_{\sst 0}\vert^2$, and both the inertial and nonlinear terms must be
considered, but at larger length scales the time derivative dominates the nonlinear term.  
At the distance $R_1$ of the closest tracers, the nonlinear term is already only $\sim 10\%$ of the inertial term and
can be ignored. We have confirmed this by noting the absence of components at frequencies of $2\omega$ in the 
power spectrum of the tracer displacements. To find the motion of the tracers we thus examine the 
solution of the unsteady Stokes equation $\rho_f\vecu_t = \Nab\cdot\vecsig$,
together with the continuity equation and boundary conditions (a) 
$\vecu = \vecv_{\sst 0}\,\expo^{\icomps\omega t}\,\hat{\vecx}$
at $r = a_0$ (particle surface), and $\vecu_{\sst 0} \rightarrow 0$ as $r \rightarrow \infty$.
The solution, first derived by Stokes \cite{Stokes1851},
can be written as ${\vecu}={\vecu}_{\sst 0}\expo^{\icomps\omega t}$, where
\begin{equation}
{\vecu}_0({\bf r}) = v_{\sst 0}\left[A(r)\vecI 
+ B(r)\,\hat{\vecr}\hat{\vecr}\right]\cdot\hat{\vecx},
\label{eq:unsteady_flow}
\end{equation}
where $\hat{\vecr}$ and $\hat{\vecx}$ are the unit vector along the radial direction and $x$-axis respectively, and
\begin{linenomath*}
\begin{subequations}
\begin{align}
A(r) &=\frac{3\alpha_{\sst 0}}{2\rho^3}
\left[(1+\rho+\rho^2)\expo^{\alpha_{\sst 0}-\rho}-1-\alpha_{\sst 0}-\frac{\alpha_{\sst 0}^2}{3}\right],
\label{eq:def_const_A1}\\
B(r) &=\frac{3\alpha_{\sst 0}}{2\rho^3}\left[3+3\alpha_{\sst 0}+\alpha_{\sst 0}^2-(3+3\rho+\rho^2)
\expo^{\alpha_{\sst 0}-\rho}\right],
\label{eq:def_const_B1}
\end{align}
\end{subequations}
\end{linenomath*}
where $\rho=(1+i)r/\delta$. 
For comparison if the sphere were moving along the $x$-axis at a constant speed 
$v_{\sst 0}$ the $\omega\to 0$ limit of \eqref{eq:unsteady_flow} has the coefficients appropriate to steady flow,
\begin{equation}
A_{\sst s} = \frac{3\alpha_{\sst 0}}{4\rho}+\frac{\alpha_{\sst 0}^3}{4\rho^3}, \ \ \ \ 
B_{\sst s} = \frac{3\alpha_{\sst 0}}{4\rho}-\frac{3\alpha_{\sst 0}^3}{4\rho^3},
\label{eq:def_const_Cstar}
\end{equation}
where these expressions are purely real since $\alpha_{\sst 0}/\rho=a_0/r$.

We now seek the motion of a probe whose equilibrium position in the oscillating flow  
is ${\bf r}=(R,\theta)$. As we are not referring to any particular tracer, we drop the 
superscript $ij$. 
The approach adopted here is a composite one, in which the Lagrangian inertial response of the tracer 
is computed from the Eulerian flow generated by the central bead.  This is valid provided 
velocity gradients at the probe scale are small, as are the oscillation amplitude relative 
to the bead radius. 
Momentum balance for a tracer
with velocity $\vecv_{\sst 1}\expo^{\icomps \omega t}$ in a fluid   
with velocity ${\vecu}_{\sst 0}\expo^{\icomps\omega t}$ takes the form \cite{Stokes1851,MazurBedeaux}
\begin{equation}
i\omega \rho_{\sst 1}\mathcal{V}_{\sst 1}\vecv_{\sst 1} = 
6\pi\eta a_{\sst 1}\!\left(\Lambda_{\sst 1}\vecu_{\sst 0} -  
\Omega_{\sst 1}\vecv_{\sst 1}\right),
\end{equation}
where $\mathcal{V}_{\sst 1} = (4/3)\pi a_{\sst 1}^3$ is the volume of the particle and 
\begin{equation}
\Lambda_{\sst 1} =  1 + \alpha_{\sst 1} + \frac{\alpha_{\sst 1}^2}{3}, \ \ \ 
\Omega_{\sst 1} =  1 + \alpha_{\sst 1} + \frac{\alpha_{\sst 1}^2}{9}, \ \ \ \ {\rm with} \ \ \ \
\alpha_{\sst 1} = (1+i)\frac{a_{\sst 1}}{\delta}.
\end{equation}

In discussing the motion of the driven particle (c.f. Eq. \ref{eq:omegazero}), we noted that the finite-frequency 
corrections to the drag law were very small; they are even smaller for tracers, whose radii
are smaller by a factor of five.  It follows that we may safely take $\Lambda_1=\Omega_1=1$,
and thus $i\omega\tau_{d_1}\vecv_{\sst 1} \simeq ({\vecu}_{\sst 0} -\vecv_{\sst 1})$, where in parallel with 
\eqref{taud} we define $\tau_{d_1}=(2\rho_1/9\rho) a_{\sst 1}^2/\nu\sim 5\times 10^{-8}\,$s.  This relaxation 
time is so short relative to the period of driven particle oscillations 
that we may assume the tracer particle velocity relaxes to that of the 
fluid instantaneously, and thus the tracer particle velocity is simply
\begin{equation}
\vecv_{\sst 1} = v_{\sst 0}\left[A(r)\vecI + B(r)\,\hat{\vecr}\hat{\vecr}\right]\cdot\hat{\vecx}\,\expo^{\icomps \omega t}.
\label{eq:dyn_resp_2}
\end{equation}
Combining this result with the response of the driven microsphere, 
the equations of motion for the tracer displacements, $\dot{\vecxi}=\vecv_{\sst 1}$, integrate to
give the tracer motion at $(R,\theta)$,
\begin{equation}
\xi_x(t) = \xi_{\sst 0}\Re\left\{
\left[A(R) + B(R)\cos^2\theta\right]\expo^{\icomps \omega t}\right\}\ \ \ \  {\rm and} \ \ \ \
\xi_y(t)=\xi_{\sst 0}\Re\left\{B(R)\cos\theta\sin\theta\,\expo^{\icomps \omega t}\right\}\,.
\label{eq:xy}
\end{equation}

While in direct comparison with experiment we utilize the full expressions in \eqref{eq:xy}, it is 
heuristically useful to simplify these results in the regime of low frequencies, when the distances $R_i$ 
of the tracers from the drive sphere are small compared to
the viscous penetration depth $\delta$. Thus expanding \eqref{eq:def_const_A1} and \eqref{eq:def_const_B1} for
$\alpha_{\sst 0},\rho \ll 1$ we find
\begin{linenomath*}
\begin{subequations}
\begin{align}
    A(R)&\simeq A_s(R)+\alpha_{\sst 0} \left(A_s(R)-1\right)+ \cdots,\\
    B(R)&\simeq B_s(R)+\alpha_{\sst 0} B_s(R) + \cdots.
\end{align}
\end{subequations}
\end{linenomath*}
Substituting into \eqref{eq:xy}, assuming as above $R/\delta\ll 1$, we observe that $A_s(R)$ and $B_s(R)$ 
are dominated by their Stokeslet contributions $3a_0/4R$, and thus
\begin{linenomath*}
\begin{equation}
\xi_x(t)\simeq \gamma\left(1+\cos^2\theta\right)\cos(\omega t+\phi_x), \ \ \ \ \
\xi_y(t)\simeq \gamma\sin\theta\cos\theta\cos(\omega t+\phi_y),
\label{lissajousxy}
\end{equation}
\end{linenomath*}
where $\gamma(R)=(3a_0/4R)\xi_{\sst 0}$, and assuming the phase shifts are small, we find
\begin{equation}
    \phi_y\simeq \frac{a_{\sst 0}}{\delta} \ \ \ \ {\rm and} \ \ \ \ \phi_x\simeq \phi_y -\frac{4R}{3\delta}\frac{1}{1+\cos^2\!\theta}.
    \label{eq:phixy}
\end{equation}
Interestingly, while $\phi_x$ varies with the polar angle, $\phi_y$ does not.

The Lissajous figures associated with \eqref{lissajousxy} are conic sections \cite{conics}, and can be rewritten as
\begin{equation}
    \frac{\sin^2\!\theta}{\left(1+\cos^2\!\theta\right)^2}\,\xi_x^2
    -\frac{2\tan\theta\cos\phi}{1+\cos^2\!\theta}\, \xi_x\xi_y
    +\frac{1}{\cos^2\!\theta}\, \xi_y^2-\gamma^2\sin^2\!\theta\sin^2\!\phi=0,
    \label{eq:lissajous_xy}
\end{equation}
where the phase shift difference $\phi=\phi_y-\phi_x$ is
\begin{equation}
\phi(R,\theta) \simeq \frac{4}{3}\frac{R}{\delta}\frac{1}{1+\cos^2\theta}.
\label{eq:phi}
\end{equation}

Equation \ref{eq:lissajous_xy} is in the standard form ${\cal A}\xi_x^2+{\cal B}\xi_x\xi_y
+{\cal C}\xi_y^2+{\cal F}=0$ of conic 
sections, which are in this cases ellipses since 
${\cal D}\equiv {\cal B}^2-4{\cal AC}=-4[\tan\theta\sin\phi/(1+\cos^2\!\theta)]^2<0$.  A standard analysis shows that the
major axis of the ellipse is tilted with respect to the $x$-axis by an angle $\psi$ satisfying $\tan 2\psi={\cal B}/({\cal A}-{\cal C})$.
As $\psi$ varies with $\cos\phi$, corrections to the $\phi=0$ limit are ${\cal O}((R/\delta)^2)$, so $\psi$ is well-approximated by the tilt angle of a steady stokeslet,
\begin{equation}
\psi_s=\tan^{-1}\left(\frac{\sin\theta\cos\theta}{1+\cos^2\!\theta}\right).
\label{tiltangle}
\end{equation} 
The fundamental signature of unsteadiness in the present experiment is the elliptical form of the tracer orbits.  From the
general expression for the semimajor and semiminor axes of ellipses,
\begin{equation}
    a^2,b^2=\frac{2{\cal F}}{{\cal D}}\left\{{\cal A}+{\cal C} \pm \left[\left({\cal A}+{\cal C}\right)^2+{\cal D}\right]^{1/2} \right\},
\end{equation}
where the $+$ ($-$) sign refers to $a$ ($b$), 
we obtain the remarkably simple asymptotic results,
\begin{linenomath*}
\begin{equation}
    \frac{a}{\xi_{\sst 0}}= \frac{3a_{\sst 0}}{4R}\frac{1+\cos^2\!\theta}{\cos\psi_s}+\cdots, \ \ \ \ {\rm and} \ \ \ \
    \frac{b}{\xi_{\sst 0}}= \frac{a_{\sst 0}}{\delta}\vert\sin\psi_s\vert + \cdots.
    \label{eq:ellipse_axes}
\end{equation}
\end{linenomath*}
Equations \ref{eq:ellipse_axes} are the heuristic results we sought.  They show that to leading order at low
frequencies the semimajor axis $a$ is given simply by the motion of the 
driven particle, projected to its position via the stokeslet contribution, while the semiminor axis $b$ is 
nonzero only to the extent that
the viscous penetration length itself is not infinite.  The aspect ratios of the ellipses simply reflect the phase shift;
$b/a=(\phi/2)\vert\sin2\psi_s\vert$, and in the steady limit the ellipses degenerate into lines.
For later reference we note if we include the leading unsteady corrections, then the normalized 
$x$-component of the tracer displacements can be written as
\begin{equation}
    \chi \equiv \frac{\max\{\xi_{x}\}}{\xi_{\sst 0}}= \chi_s-\frac{a_{\sst 0}}{\delta}+\cdots
    \,, \ \ \ \ {\rm with} \ \ \ \ 
    \chi_s=\frac{3a_{\sst 0}}{4R}\left(1+\cos^2\!\theta\right)\,.
    \label{eq:chi}
\end{equation}

\begin{figure}[t]
\begin{center}
\scalebox{0.98}{\includegraphics{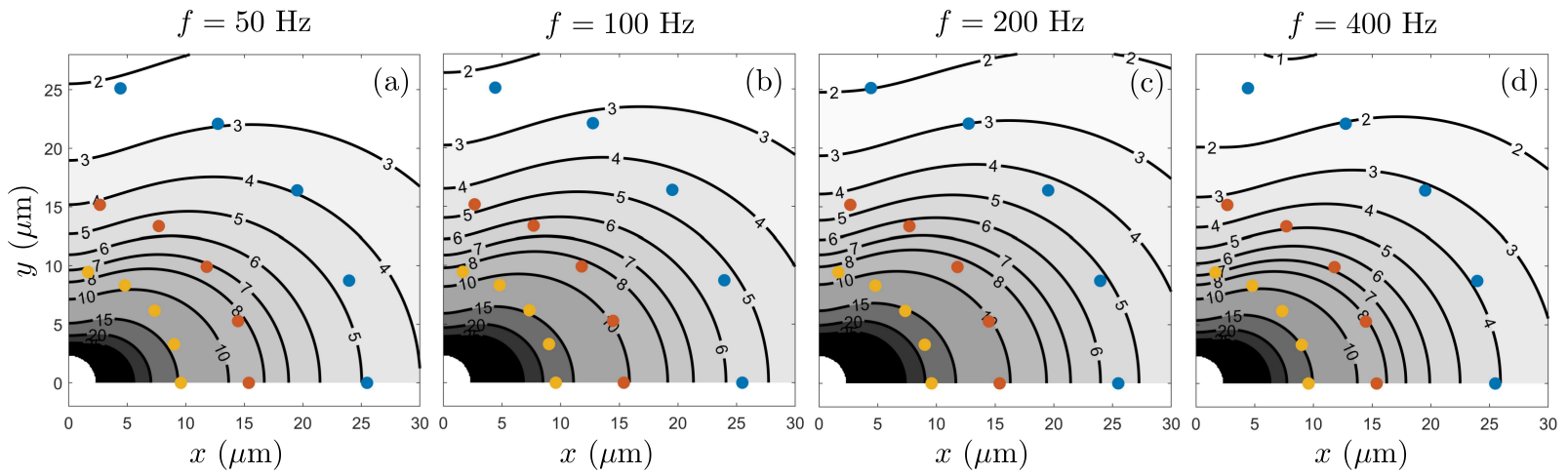}}
\caption{Significance of thermal fluctuations.  (a)-(d) Contour plots of the $\sqrt{Pe}$ 
\eqref{eq:peclet} in physical space, with the locations of the probe microspheres superimposed, at the 
four driving frequencies.  
\label{fig:peclet}}
\end{center}
\end{figure}

\begin{figure}[b]
\begin{center}
\scalebox{0.95}{\includegraphics{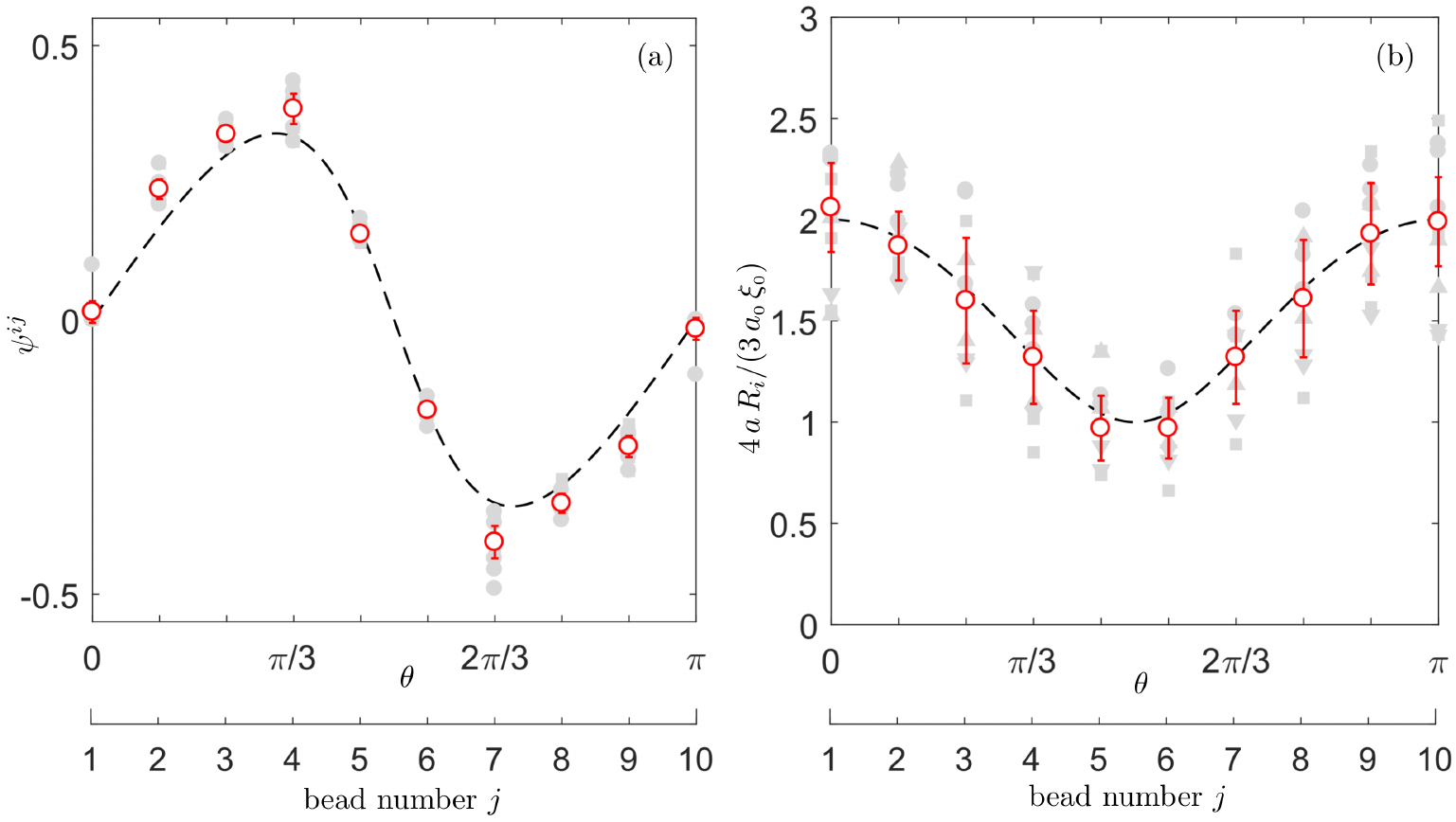}}
\caption{Orientation and size of elliptical tracer trajectories.  (a) Tilt angle of the major axis of elliptical
Lissajous figures as a function of mean angular position relative to driving axis. Gray symbols represent the individual
data points at the four experimental frequencies ($f=50$, $100$, $200$, and $400$ Hz) and three radii ($R=9.5$, $15.8$, and 
$25.4\,$ $\mu$m)
at a given angular position $\theta_i$.  Open red circles denote the mean values of each of those measurements at a given 
$\theta$ and their standard deviations.  Dashed line is the steady tilt angle \eqref{tiltangle}. (b) As in (a), but
for the semimajor axis of the ellipses.  Dashed line is the low-frequency limit \eqref{eq:ellipse_axes}.}
\label{fig:ellipses}
\end{center}
\end{figure}

Before analyzing the tracer trajectories in detail, we use the  
particle trajectories to quantify the 
competition between the deterministic forcing of the tracer particles 
and thermally-driven Brownian motion.  As in previous discussions of this 
issue \cite{Leptos}, a useful metric with which to assess these effects is 
the ratio of the maximum deterministic displacement due to the oscillating fluid (the major axis of the ellipse) to the 
average Brownian displacement over 
half an oscillation period.  This is essentially the square root of a P{\'e}clet 
number,
\begin{equation}
    \sqrt{Pe}=\frac{2a}{(2\pi D_{\sst 1}/\omega)^{1/2}}\simeq \frac{3}{2\sqrt{\pi}}\left(\frac{a_{\sst 0}}{R}\right)
    \left(\frac{\xi_{\sst 0}}{\delta}\right)\sqrt{\frac{\nu}{D_{\sst 1}}}\frac{1+\cos^2\!\theta}{\cos\psi_s}.
    \label{eq:peclet}
\end{equation}
Here, $D_{\sst 1} = k_{\sst B} T /6 \pi \eta a_1\sim 0.4\, \mu$m$^2$/s (Table 1) is the diffusion constant of the tracers, 
with $k_{\sst B}$ the
Boltzmann constant and $T = 298$ K the absolute temperature.  The last relation in \eqref{eq:peclet} is 
obtained using the asymptotic results above and displayed to emphasize it is the product of three
dimensionless ratios. The factor $\nu/D_{\sst 1}$ is a Schmidt number 
$Sc$ for the tracer particles and is very large ($\sqrt{Sc}\simeq 1.6\times 10^{3}$), but its
contribution to $Pe$ is attenuated by the two small factors $a_{\sst 0}/R$ and $\xi_{\sst 0}/\delta$,
each on the order of $0.1$.  The frequency dependence of $\sqrt{Pe}$ is relatively weak by virtue of the counteracting
trends of $\xi_{\sst 0}\sim \omega^{-1}$ and $\delta\sim \omega^{-1/2}$.
Contour plots of \eqref{eq:peclet} in the first quadrant of physical space where the tracers reside are shown in 
Figure \ref{fig:peclet}, in which the semimajor axis has been computed with the full unsteady velocity field 
given in \eqref{eq:unsteady_flow}, \eqref{eq:def_const_A1} and
\eqref{eq:def_const_B1}. From these results we see that advective contributions dominate diffusion 
($Pe>1$) at all 
frequencies for the innermost spheres, while the two become comparable for the outermost spheres, consistent
with the qualitative appearance of the trajectories.

\begin{figure}[t]
\begin{center}
\scalebox{1.0}{\includegraphics{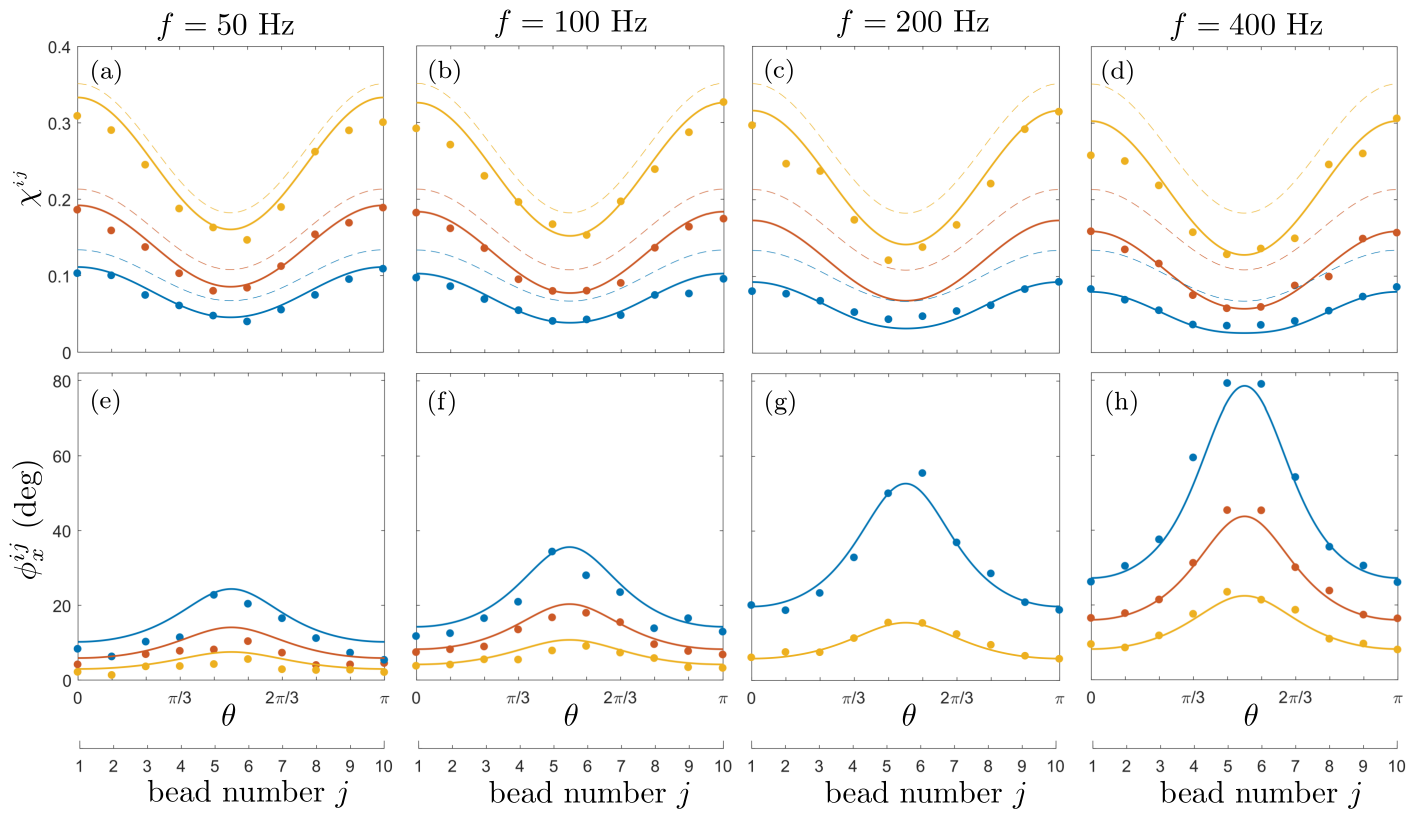}}
\caption{Dynamics of the passive tracers. (a-d) Normalized maximum displacements (solid circles), and their 
theoretical counterpart $\chi$ (solid lines). Dashed line show displacements calculated within steady Stokes equation.
(e-h) Phase shifts between responses of the probes
and the harmonic motion of driving bead (solid circles), and theoretical counterpart (solid lines).}
\label{fig:ampl_phase}
\end{center}
\end{figure}

A first test of the theoretical analysis of the trajectories involves plotting the ellipses from 
Eq.~\ref{eq:lissajous_xy} in the $x,y$ plane. These are shown in Figs.~\ref{fig:traj}(e-h), magnified by a 
factor 4 to be consistent with all the plots of the figure.  In addition, the boxed portion of Fig.~\ref{fig:traj}(f) 
is expanded in (j). In comparison, Fig.~\ref{fig:traj}(i) displays the average oscillation of a few tracers at the 
same locations, from experiments, showing a good match with the theory.  To suppress the effect of fluctuations on 
single oscillations, all the cycles were averaged into a composite, cyclic $x,y$ path for each tracer. Each is 
obtained by computing the average $x$- and $y$-oscillations as a histogram with $2\pi f_s / \omega$ 
bins where $f_s$ is the sampling frequency; these data are accumulated in the bins using the time 
$t \mod (2\pi / \omega)$.
Alternatively, Fig.~\ref{fig:ellipses} shows two basic geometrical features of the 
elliptical tracer trajectories, their orientation and major axis, each expected to be dominated
by their steady contributions.  The orientation angle in Fig. \ref{fig:ellipses}(a) agrees well with the 
steady angle $\psi_s$ in \eqref{eq:ellipse_axes}, and the semi-major axis (Fig. \ref{fig:ellipses}(a)) is likewise 
well described by the leading order relation in \eqref{tiltangle}.

Focusing on the displacements along the same ($x$) axis as the driven microsphere, Fig. \ref{fig:ampl_phase} 
summarizes the results for the amplitudes and phase shifts of the 
tracers. In (a-d) we plot the normalized 
component of the displacement as defined in \eqref{eq:chi},
\begin{equation}
    \chi^{\sst ij} = \frac{\max\{\xi_{x}^{\sst{ij}}\}}{\xi_{\sst 0}}
    \label{eq:chiij}
\end{equation}
for experiments (symbols) and theory (lines).  In the experiments, the 
relative amplitude and phase compared to the driven bead are measured from the fast Fourier transform (FFT) of 
the $x(t)$ data of each tracer, by identifying the amplitude (respectively, the phase) from the peaks 
in the magnitude (phase) plots of the FFT for the tracers and dividing by the magnitude (or subtracting the 
phase) from the FFT of the driven bead. 
At any given frequency the agreement between the data and the steady theory (shown by dashed lines) is best 
for those tracers closest to the driven particle and progressively decreases for more distant probes, while 
at any given radius the agreement with the steady theory worsens at higher frequencies.  
For example, the deterministic component of the displacement is overestimated by about
20\% for $f = 50$ Hz, and up to 100\% (for $f = 400$~Hz) for the most remote ones - i.e at a distance 
$R_{\sst 3}$ from the origin. Both of these trends
are fully consistent with the relevant measure of unsteadiness being $R/\delta$. In Fig. \ref{fig:ampl_phase}b
we show the the experimental phase shift $\phi_x^{\sst ij}$ between the tracers and the active particle.
The magnitude and angular dependence are both accurately captured by the unsteady theory.
At the very highest frequency used, the phase shifts of the most distant probes located close to the
$y$-axis \textemdash at $(R_{\sst 3},\theta_{\sst 5})$ and $(R_{\sst 3},\theta_{\sst 6})$\textemdash are 
very large; the probes are almost in quadrature with the forcing. 
We see from these results that despite a large displacement of the central bead (which is of the same order 
as its radius) and a direct use of the Eulerian form of
the viscous unsteady flow $\vecu_{\sst 0}$, the agreement between theory and experiment is very good,
with a maximum relative error of $4\%$ overall.

\begin{figure}
\begin{center}
\scalebox{0.99}{\includegraphics{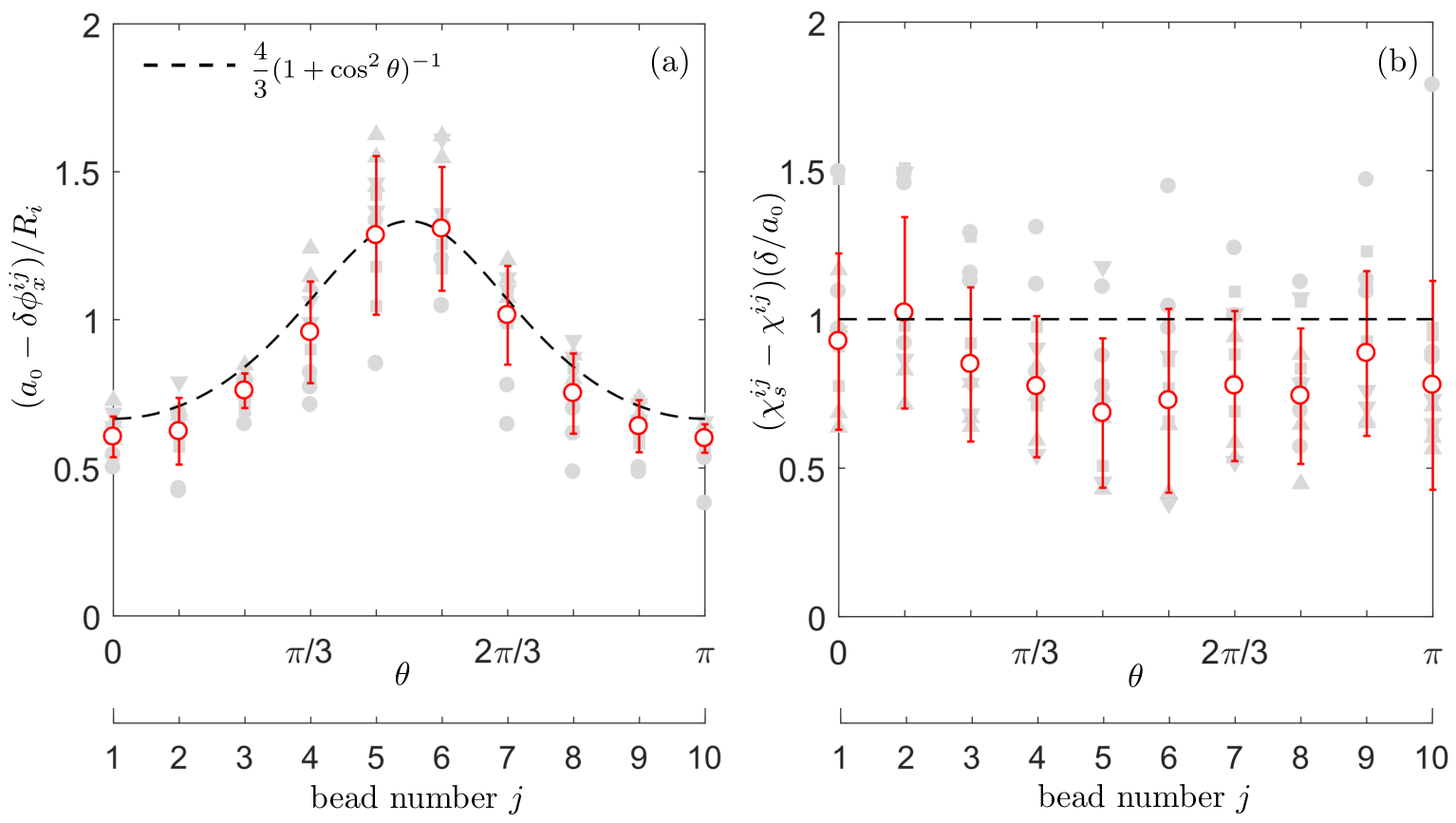}}
\caption{Test of the low-frequency scaling. (a) Rescaled phase shift as a function of angle, using same symbol motif as 
in Fig. \ref{fig:ellipses}. Dashed line is the scaling result \eqref{eq:phi_scaling}. 
(b) As in (a), but for the oscillation amplitude, with dashed line from \eqref{eq:phi_scaling} to leading order.  There are no adjustable parameters
in (a) or (b).\label{fig:collapse}}
\end{center}
\end{figure}

As a final test of the unsteady theory, we ask whether the data in Fig. \ref{fig:ampl_phase} are consistent with the predicted
leading-order low-frequency limits in the sense of a data collapse.  Focusing on the same $x$-component of the phase shift and amplitude,
the analysis in \eqref{eq:phixy} and \eqref{eq:phi} can be written the scaling forms for the phase and amplitude,
\begin{equation}
    \frac{a_{\sst 0}-\delta\phi_x}{R} = \frac{4}{3}\frac{1}{1+\cos^2\!\theta}\,, \ \ \ \ {\rm and} \ \ \ \ 
    \frac{\delta}{a_{\sst 0}}\left(\chi_s-\chi\right)= 1+\cdots\,.
    \label{eq:phi_scaling}
\end{equation}
Figure \ref{fig:collapse} shows good agreement in both cases, especially for the phase shift.

\section{Discussion \label{sec:discuss_concl}}

We have shown here how optical trapping and particle-tracking 
techniques allow for a precise microscale test of the 
theory of unsteady Stokes flows.  At the scale of colloidal particles, and with oscillation frequencies in the range
found in biological systems, the simplifications arising from lack of inertial effects on particle motion and
corrections to the Stokes drag law allow for a simple and compact picture of the particle orbits.
The regime of sizes and frequencies 
explored is also such that Brownian motion makes only a modest contribution to the tracer dynamics, with an effective 
P{\'e}clet number generally exceeding unity. Our experimental results show 
that tracer particles move on simple elliptical orbits even in a regime with very large phase shifts, 
in quantitative accord with a low-frequency analysis.  These experimental observations would be difficult to
reproduce by conventional particle imaging techniques which are based on obtaining an Eulerian velocity map from 
correlation functions of small tracer displacements.

As outlined in the introduction, one clear motivation for the present study is provided by the evidence
that unsteady effects are present during the collective beating of eukaryotic cilia and flagella.  It is an open 
question as to whether these effects 
actually control synchronization.  Two features of the present work will likely bear on this issue; the
angular dependence of the phase shift and the elliptical orbits themselves.  The former dictates the
strength of the lateral coupling between cilia along a tissue, while the latter represents vorticity
created by the driven particle that may be relevant to wave propagation. 
It is important to note 
that in the many cases in which metachronal waves occur there is a nearby underlying
no-slip surface \textemdash the cell wall of a ciliate, 
or the tissue surface of an ciliated epithelium \textemdash whose presence can not be ignored.
Indeed, a recent study of synchronization in arrays of  
``colloidal oscillators" \cite{MW9}, microspheres moved along periodic orbits by optical traps, show that
surface proximity can profoundly affect the collective dynamics that they exhibit.  Thus, a natural next step
is the study of model unsteady flows near no-slip surfaces \cite{unsteady_image}.  

\section{acknowledgements}

This work was supported in part by ERC Consolidator grant 682754 (EL), ERC PoC grant CellsBox (PC and JK),
Wellcome Trust Investigator Award  207510/Z/17/Z, Established Career Fellowship EP/M017982/1 from the
Engineering and Physical Sciences Research Council, and the Marine Microbiology Initiative of the 
Gordon and Betty Moore Foundation, Grant 7523 (REG).


\begin{thebibliography}{99}
\bibitem{missingdata} The data for the red tracers ($R_2,f=200$ Hz) in 
Figs. \ref{fig:traj}(c,g) are not available due to unforeseen data corruption on the hard-drive.  
The missing trajectories do not affect the overall trend and results presented here.
\end{thebibliography}
\end{document}